# Photon-counting Brillouin optical time-domain reflectometry based on up-conversion detector and fiber Fabry-Perot scanning interferometer

Haiyun Xia,[1,3] Mingjia Shangguan,[1] Guoliang Shentu,[2,4] Chong Wang,[1] Jiawei Qiu,[1] Xiuxiu Xia,[2,4] Chao Chen,[2,4] Mingyang Zheng,[5] Xiuping Xie,[5] Qiang Zhang,[2,4,6] Xiankang Dou,[1,7] and Jianwei Pan[2,4,8]

[1]*CAS Key Laboratory of Geospace Environment, University of Science and Technology of China, Hefei, 230026, China*
[2]*Shanghai Branch, National Laboratory for Physical Sciences at Microscale and Department of Modern Physics, University of Science and Technology of China, Shanghai, 201315, China*
[3]*Collaborative Innovation Center of Astronautical Science and Technology, HIT, Harbin 150001, China*
[4]*Synergetic Innovation Center of Quantum Information and Quantum Physics, USTC, Hefei 230026, China*
[5]*Shandong Institute of Quantum Science and Technology Co., Ltd, Jinan, Shandong 250101, P. R. China*
[6]*e-mail:qiangzh@ustc.edu.cn,*
[7]*e-mail: dou@ustc.edu.cn*
[8]*e-mail:pan@ustc.edu.cn*

A direct-detection Brillouin optical time-domain reflectometry (BOTDR) is proposed and demonstrated by using an up-conversion single-photon detector and a fiber Fabry-Perot scanning interferometer (FFP-SI). Taking advantage of high signal-to-noise ratio of the detector and high spectrum resolution of the FFP-SI, the Brillouin spectrum along a polarization maintaining fiber (PMF) is recorded on a multiscaler with a small data size directly. In contrast with conventional BOTDR adopting coherent detection, photon-counting BOTDR is simpler in structure and easier in data processing. In the demonstration experiment, characteristic parameters of the Brillouin spectrum including its power, spectral width and frequency center are analyzed simultaneously along a 10 km PMF at different temperature and stain conditions. © 2014 Optical Society of America

For over two decades, distributed optical fiber sensors based on Brillouin scattering has attracted intensive interests, because it not only inherits features from fiber sensors, such as durability, stability, small size and immunity to environmental EM perturbations, but also permits simultaneous temperature and stain detection in large infrastructures, aerospace industry, and geotechnical engineering [1, 2].

These sensors can be divided into two types. In the first type called Brillouin optical time-domain analyzer (BOTDA), a pulsed pump light is launched at one end of an optical fiber and interacts with the counterpropagating CW probe light injected from the other end. The probe light at Stokes (anti-Stokes) frequency is amplified (attenuated) by the pump light through the Brillouin gain (loss) process. By measuring the time dependent CW signal over a wide range of frequency difference between the pump and probe, the Brillouin spectrum at each fiber location could be analyzed. To increase the sensing length, the Brillouin loss regime is investigated rather than the gain regime [3]. In the second type called Brillouin optical time-domain reflectometer (BOTDR), a pulsed light is launched into only one end of an optical fiber, and Brillouin backscattering is mixed with a reference light and homodyne detection is performed at the optical coherent receiver. The Brillouin spectrum at any position along the optical fiber is measured by scanning the frequency difference between the local oscillation and the probe pulse [4]. The Stokes Brillouin signal is measured in BOTDR, since its power is generally higher than the anti-Stokes Brillouin signal [5].

In some special circumstances for example in an oil well, where light is only allowed to input through one end, BOTDR is the unique choice. However, in the BOTDR, homodyne beating of backscattering with local oscillation is recorded using an analog to digital converter (ADC). High sampling frequency is required to improve the spatial resolution, resulting heavy burden in post-processing the big data at the sacrifice of temporal resolution.

Here, in this letter, a new photon-counting BOTDR is proposed, where the Brillouin spectra along a polarization maintaining fiber (PMF) is reconstructed with the backscattering signals passing through a fiber Fabry-Perot scanning interferometer (FFP-SI) directly. The backscattering signals are detected by an upconversion single photon detector and recorded on a multiscaler via photon-counting acquisition with a small data volume comparing to its conventional counterpart.

Actually, photon-counting OTDR based on Rayleigh backscattering has already received increasing attention [6-11]. Rayleigh scattering is around 20 times stronger than the Brillouin scattering. Thus, analyzing the Brillouin spectra along a PMF is a great challenge.

The structure for our photon-counting BOTDR is illustrated in Fig. 1. In the demonstration experiment, the light source is designed for an aerosol Lidar at 1548.1 nm [12]. The continuous wave from a seed laser (Keyopsys, PEFL-EOLA) is chopped into pulse train after passing through an acousto-optic modulator (AOM) (Gooch & Housego, T-M080). The AOM is driven by an arbitrary waveform generator (Agilent, 33250A), which determines the shape of the laser pulse and its repetition rate. The weak laser pulse is fed to a polarization maintaining EDFA (Keyopsys, PEFA-EOLA), which can deliver pulse train with pulse energy up to 110 μJ. In this work the peak power of the laser pulse is set to 0.1 W by tuning the pump current of the EDFA.

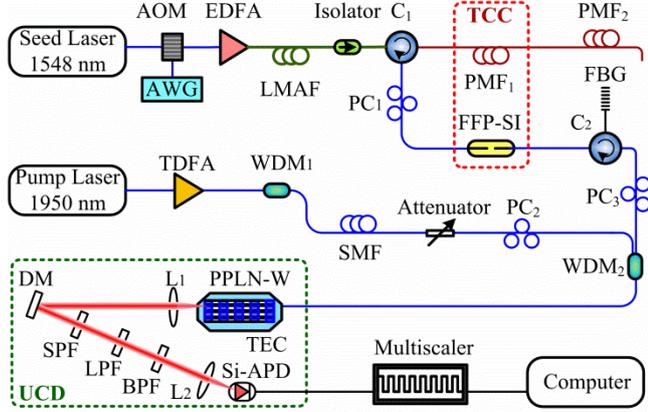

Fig. 1. System layout of the photon-counting Brillouin optical time-domain reflectometry. AOM, acousto-optic modulator; AWG, arbitrary waveform generator; EDFA, erbium doped fiber amplifier; LMAF, large mode area fiber; C, circulator; PC, polarization controller; TCC, temperature controled chamber; FFP-SI, fiber Fabry-Perot scanning interferometer; FBG, fiber Bragg grating; TDFA, thulium doped fiber amplifier; WDM, wavelength division multiplexer; SMF, single mode fiber; TEC, thermoelectric cooler; MMF, multimode fiber; DM, dichroic mirror; SPF, short-pass filter; LPF, long-pass filter; BPF, band-pass filter.

The $PMF_1$ and the FFP-SI are put in a chamber with temperature fluctuation controlled within $\pm 0.01\ °C$. Another $PMF_2$ is used as the sensing fiber. The backscattering signal is coupled into the FFP-SI via a circulator ($C_1$). An ultra-narrow fiber Bragg grating (FBG) in combined with another circulator ($C_2$) is used to pick out the Brillouin signal from the strong Rayleigh signal.

Different single photon detectors have been exploited in OTDR experiments, including InGaAs/InP avalanche photodiode (APD) [6], up-conversion detector (UCD) [7-9], and superconducting nanowire single-photon detector (SNSPD) [10, 11]. Although InGaAs APD is commercial available, its high after-pulse probability distorts the raw data significantly and the time gating operation exhibits periodic blind zones [7]. And, the SNSPD requires liquid helium refrigeration equipment to achieve high efficiency and low noise. Considering the photon-counting detector is the workhorse device in field operation, an UCD based on periodically poled Lithium niobate waveguide (PPLN-W) is designed and manufactured in this work.

The continuous wave from the pump laser at 1950 nm is followed by a thulium-doped fiber amplifier, both manufactured by AdValue Photonics (Tucson, AZ). The residual ASE noise is suppressed by using a 1.55/1.95-μm wavelength division multiplexer ($WDM_1$). The Brillouin backscattering and the pump laser are coupled into a self-made PPLN-W (manufactured at the Shandong Institute of Quantum Science and Technology) via the second $WDM_2$. Optimized quasi-phase matching condition is achieved by tuning the temperature of the PPLN-W with a thermoelectric cooler. The backscattering photons at 1548 nm are converted into sum-frequency photons at 863 nm and then picked out from the pump and spurious noise by using a combination of filters, including a dichroic mirror, a short-pass filter (945 nm), a long-pass filter (785 nm) and a band-pass filter (863 nm) as shown in Fig. 1. Finally, the photons at 863 nm are focused onto a Si: APD (EXCELITAS, AQRH16). The TTL signal corresponding to the photons received on the Si-APD is recorded on a multiscaler (FAST ComTec, MCS6A) and then processed in a computer. In this work, by adjusting the pump power, the quantum efficiency of the UCD is tuned to 17%, with a noise level of 700 counts per second. For reader's convenience, key parameters of the system are listed in table 1.

Table.1 Key parameters of the system

| Parameter | Value |
| --- | --- |
| Pulsed Laser | |
|     Pulse duration(ns) | 300 |
|     Peak power (W) | 0.1 |
|     Pulse repetition rate (kHz) | 8 |
| Pump Laser | |
|     Wavelength (nm) | 1950 |
|     Power (mW) | 300 |
| FFP-SI | |
|     Free Spectrum range (GHz) | 4.02 |
|     Full width at half-maximum (MHz) | 120 |
|     Insert loss (dB) | 2.25 |
| PPLN waveguide | |
|     QPM period (μm) | 20 |
|     Length (mm) | 52 |
|     FWHM (nm) | 0.3 |
|     Insert loss (dB) | 1.4 |
| Tunable fiber Bragg grating | |
|     Tuning range (nm) | ±0.24 |
|     3-dB Bandwidth (pm) | 5 |
|     Sideband suppression (dB) | 35 |
| Si: APD | |
|     Detection efficiency at 863 nm (%) | 45 |
|     Dark count (Hz) | 10 |
|     Maximum count rate (MHz) | 43 |

Free-space bulk Fabry-Perot interferometers have been used as frequency discriminators in a temperature lidar [13] and Doppler wind lidars [14, 15]. But it is hard to eliminate the parallelism error of the reflecting mirrors and the mode-dependent spectral broadening due to its illuminating condition. In this work, to conquer these challenges, a lensless, plane Fabry-Perot interferometer is adopted. The cavity is formed by two highly reflective multilayer mirrors that are deposited directly onto two carefully aligned optical fiber ends [16]. Frequency scanning of the FFP-SI is achieved by axially straining a short piece of single-mode fiber that inserted in the cavity, using a stacked piezoelectric transducer (PZT). The anti-reflection coated fiber inserted in the cavity provides appropriate confined light-guiding and eliminates secondary cavity.

The interaction between probe pulse in an optical fiber and acoustic phonons generates Brillouin backscattering.

Because the phonons decay exponentially, the Brillouin spectrum is Lorentzian in form:

$$g_B(\nu) = g_0 / \left[1 + (\nu - \nu_B)^2 / \omega_B^2 \right] \quad (1)$$

where $\omega_B$ is the half-width at half maximum (HWHM). $\nu_B$ is the Brillouin frequency shift, and $g_0$ is the Brillouin gain coefficient. Within a limited dynamic range, $\omega_B$ and $\nu_B$ are demonstrated as linear function of strain and temperature.

The FFP-SI is fabricated with single-mode fiber with a divergence negligible in the cavity, thus its transmission is approximated to a normalized Lorentzian function [17]:

$$h(\nu) = 1 / \left[1 + \nu^2 / \omega_{FPI}^2 \right] \quad (2)$$

where $\omega_{FPI}$ is the HWHM of the transfer function. Finally, the transmission function of the Brillouin signal through the FFP-SI is the convolution of Eq. (1) and Eq. (2), yields,

$$T(\nu) = g_0 / \left[1 + (\nu - \nu_B)^2 / (\omega_{FPI} + \omega_B)^2 \right] \quad (3)$$

Eq. (3) indicates that, the power fluctuation, Brillouin frequency shift and bandwidth can be retrieved easily once the transmission curve is measured.

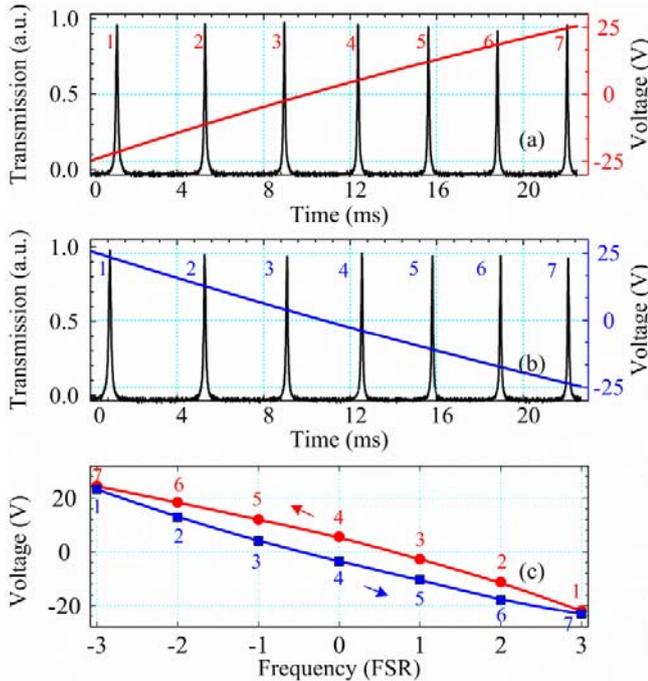

Fig. 2. Calibration of the fiber Fabry-Perot interferometer: transmission of the laser pulse through the FFPI recorded on an oscilloscope as function of the (a) increasing voltage and (b)the decreasing voltage. (c) Measured hysteresis loop of the PZT.

One problem must be considered before the detection is the hysteresis introduced by the PZT. Fortunately, it is a repeatable phenomenon, which can be calibrated in the initialization process. As shown in Fig. 2, by changing the voltage fed to the PZT, the transmission function over 7 interference orders are recorded using an oscilloscope on the up-cycle (increasing voltage) and on the down-cycle (decreasing voltage). The voltage values corresponding to transmission peaks are searched and tagged with the number of the interference order, as shown in Fig. 2.

In the frequency domain, the transmission of an FPI is periodic with a constant free spectral range (FSR). In the calibration, a C-band light from an amplified spontaneous emission (ASE) source is fed to the FFP-SI, and the FSR is read out to be 4.02 GHz directly from an optical spectrum analyzer (YOKOGAWA, AQ6370C). Then, the peak voltage values can be mapping into the frequency domain according to the number of interference order. Data fitting of the peak values on the up cycle and down cycle to 3-order polynomial functions, respectively, yields the hysteresis loop of the PZT, as shown in Fig. 2(c). So, the inherently non-linear cavity scanning due to hysteresis can be compensated in the data processing. Note that, accurate frequency analysis can only be done on either up or down cycle.

As an example, cavity scanning of the FFP-SI is performed near the zero voltage on the up cycle and then mapping into the frequency domain using the calibrated hysteresis loop. The measured transmission and its Lorentzian fitting results are shown in Fig. 3.

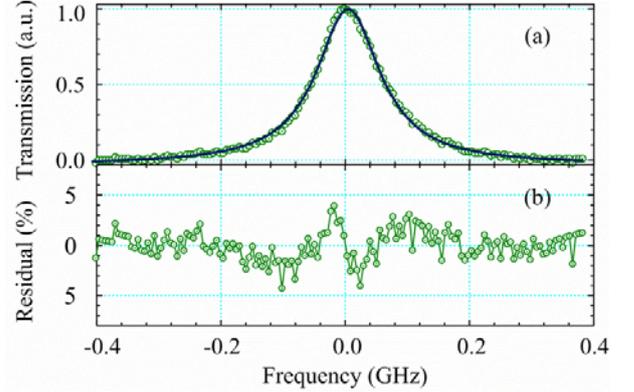

Fig. 3. Nonlinear fitting of the transmission to the Lorentz function (a) and residual (b) in % with respect to the peak.

Thanks to the ultra-narrow band and high sideband suppression of the temperature tunable FBG, we can pick out either the Stokes Brillouin spectra or the anti-Stokes Brillouin spectra for detection purpose. In the first demonstration experiment, the Stokes Brillouin spectra are investigated. An unstrained coil of $PMF_1$ (3 km) is put in the temperature controlled chamber at 19.7 °$C$, and the other coil of unstrained $PMF_2$ (9.1 km) is put at room temperature of 24.4 °$C$. The spectra along the whole fiber are scanned over 40 steps simultaneously with a sampling offset of 15 MHz/step and integration time of 1 second/step. The range resolution is 60 m, which is determined by the duration of the probe pulse. As shown in Fig. 4(a), the Brillouin frequency and bandwidth vary with the temperature distribution.

In the second demonstration, the FBG is adjusted to pick out the anti-Stokes Brillouin spectra. Another $PMF_1$ of 300 m is twined with strain and the chamber temperature is set to 32.6 °$C$, and the unstrained $PMF_2$ is still at 24.4 °$C$. Obviously, the large frequency shift between the spectra of $PMF_1$ and $PMF_2$ is beyond the contribution of temperature, as shown in Fig. 4(b). Fig. 5 shows the spectra details under different conditions and their Lorentzian fitting results.

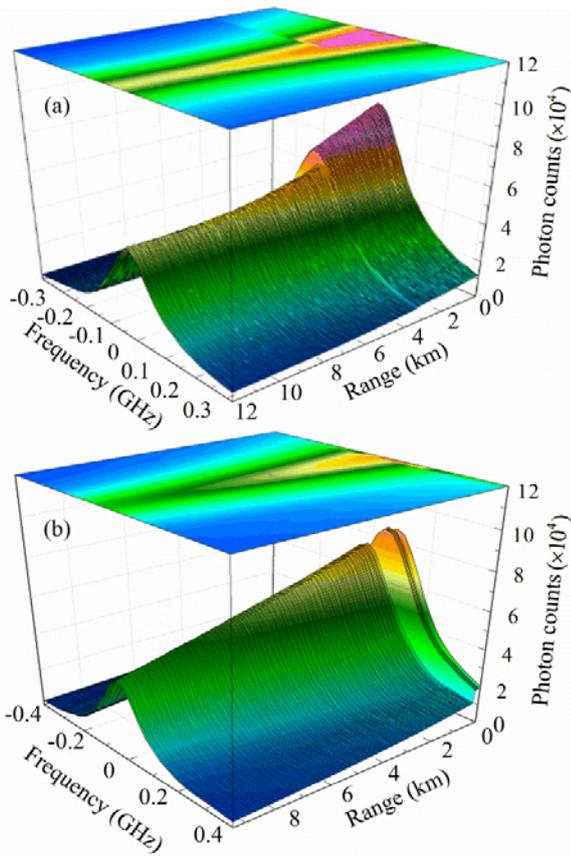

Fig. 4. (a) Stokes Brillouin Spectra of unstrained panda fibers: PMF$_1$ (3km) at 19.7 °C and PMF$_2$ (9.1km) at 24.6 °C . (b) Anti-Stokes Brillouin spectra of strained PMF$_1$ (300m) at 32.6 °C and unstrained PMF$_2$ (9.1km) at 24.4 °C .

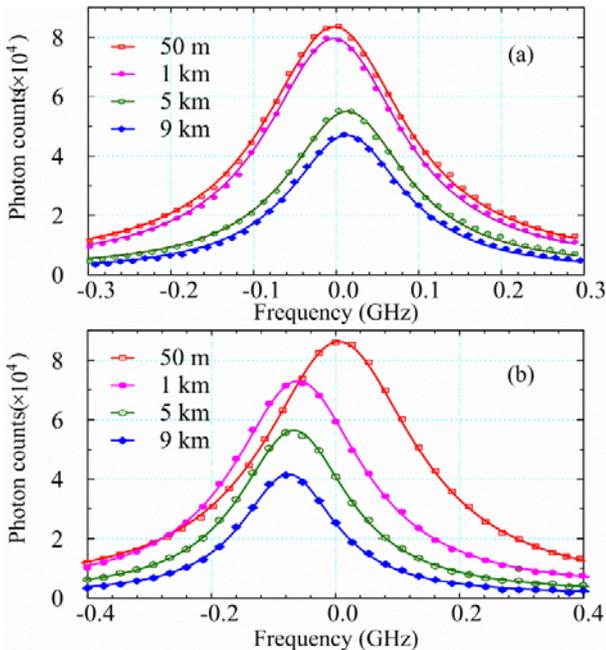

Fig. 5. Nonlinear fitting of the raw data to the Lorentz function at different ranges. (a) Fitting results when the Stokes Brillouin Spectra are scanned (b) Fitting results when the anti-Stokes Brillouin spectra are scanned.

In conclusion, for the first time, we proposed and demonstrated a simple photon-counting BOTDR based on up-conversion detector and fiber Fabry-Perot scanning interferometer. The Brillouin frequency, power and bandwidth can be retrieved along a polarization fiber for simultaneous detection of temperature and strain distributions.

This work has been supported by the National Basic Research Programm (2011CB921300, 2013CB336800), the NNSF of China, and the CAS.